\documentclass{article}

\usepackage[a4paper, margin=1.0in]{geometry}
\usepackage[utf8]{inputenc}
\usepackage[dvipsnames,table,xcdraw]{xcolor}
\usepackage[affil-it]{authblk}
\usepackage[misc]{ifsym}
\usepackage{multirow}

\usepackage{appendix,cite,verbatim,listings,bm,braket,enumitem, kantlipsum,geometry,graphicx,multicol,mathtools,csquotes,amsmath,mathtools,url,subfiles,qcircuit,amsthm,amssymb,fdsymbol,rotating,caption,hyperref,fontawesome,xcolor,arydshln,ulem,subcaption,caption, amsfonts}
\usepackage{tikz}
\usetikzlibrary{arrows.meta,positioning,fit,calc}
\usepackage{booktabs}
\usepackage{tabularx} 

\begin{document}

\title{\textbf{\LARGE From Characterization To Construction: \\ \large Generative Quantum Circuit Synthesis from Gate Set Tomography Data}}

\author[1,2]{King Yiu Yu}
\author[1,2]{Aritra Sarkar}
\author[1]{Erbing Hua}
\author[2,3]{\\Maximilian Rimbach-Russ}
\author[1,2]{Ryoichi Ishihara}
\author[1,2]{Sebastian Feld}
\affil[1]{Quantum and Computer Engineering, Delft University of Technology, The Netherlands}
\affil[2]{QuTech, Delft University of Technology, The Netherlands}
\affil[3]{Kavli Institute of Nanoscience Delft, Delft University of Technology, The Netherlands}
\date{}
\maketitle

\begin{abstract}
     High-fidelity circuit execution on noisy intermediate-scale quantum (NISQ) devices is often bottlenecked by traditional compilation pipelines that disregard complex, correlated noise. To address this, this methodology article proposes a quantum machine learning control (QMLC) framework for generative quantum circuit synthesis from gate-set tomography (GST) data that bypasses the traditional two-step pipeline of characterizing native quantum gates via GST followed by unitary decomposition algorithms. Instead, our idea is to directly learn a generative concept space from GST data which enables conditional synthesis of quantum circuits that produce a desired output distribution. Our approach tokenizes GST germ circuits and embeds them into a structured latent space using a curriculum-learning-motivated strategy, starting with short circuits and progressively incorporating longer ones with diverse output statistics. The embedded sequences are processed by a set-vision transformer with permutation-invariant pooling, producing k-seed vectors that represent the learned concept space of the quantum device. Aggregating data across multiple circuits makes this latent representation inherently ``context-aware'', capturing the shared physical noise environment (such as crosstalk and drift) that isolated gate metrics miss. We propose an unconditional diffusion model to sample from the concept space. During inference, a user provides a target measurement distribution, and the model generates a corresponding circuit. To ensure fidelity and robustness, the output is denoised using a diffusion model that operates on the target conditional covariance matrix. This end-to-end framework make steps towards context-aware, hardware-native circuit synthesis directly from raw GST data, which offers a new paradigm for integrating quantum control and compilation. This QMLC framework is particularly suited for near-term quantum devices with complex calibration procedures.
\end{abstract}

\section{Introduction}

The rapid development of quantum hardware has catalyzed the need for efficient hardware-aware circuit compilation strategies~\cite{campbell2023superstaq, sarkar2026yaqq, stano2022review, tan2024compiling, harrigan2024expressing}. 
Quantum processors are heterogeneous owing to calibration drift, cross-talk, and gate imperfections, leading to a growing demand for compilation and control techniques that are grounded in empirical online characterization of the devices~\cite{zwolak2023colloquium, katiraee2025unified}. 
Quantum process and gate-set tomography (GST)~\cite{nielsen2021gate}, and derivatives of randomized benchmarking~\cite{knill2008randomized, helsen2022general, huang2020predicting, helsen2023shadow} are used to characterize the native gates of a quantum device with high precision, providing complete and self-consistent models of gate operations, SPAM errors, and measurement maps. 
However, leveraging GST data in downstream tasks remains limited: it is typically used to calibrate or correct gate operations post hoc, and circuit design is still performed independently using standard decomposition techniques~\cite{dawson2005solovay, ross2014optimal}. 
Thus, traditional quantum compilation workflows assume idealized gate libraries, which often result in circuits that perform poorly when deployed on real hardware. 
This decoupling between characterization and compilation misses the opportunity to leverage models grounded in hardware behavior.

\begin{figure}[htb]
    \centering
    
    \includegraphics[width=1.0\linewidth,trim={19cm 0 0 4cm},clip]{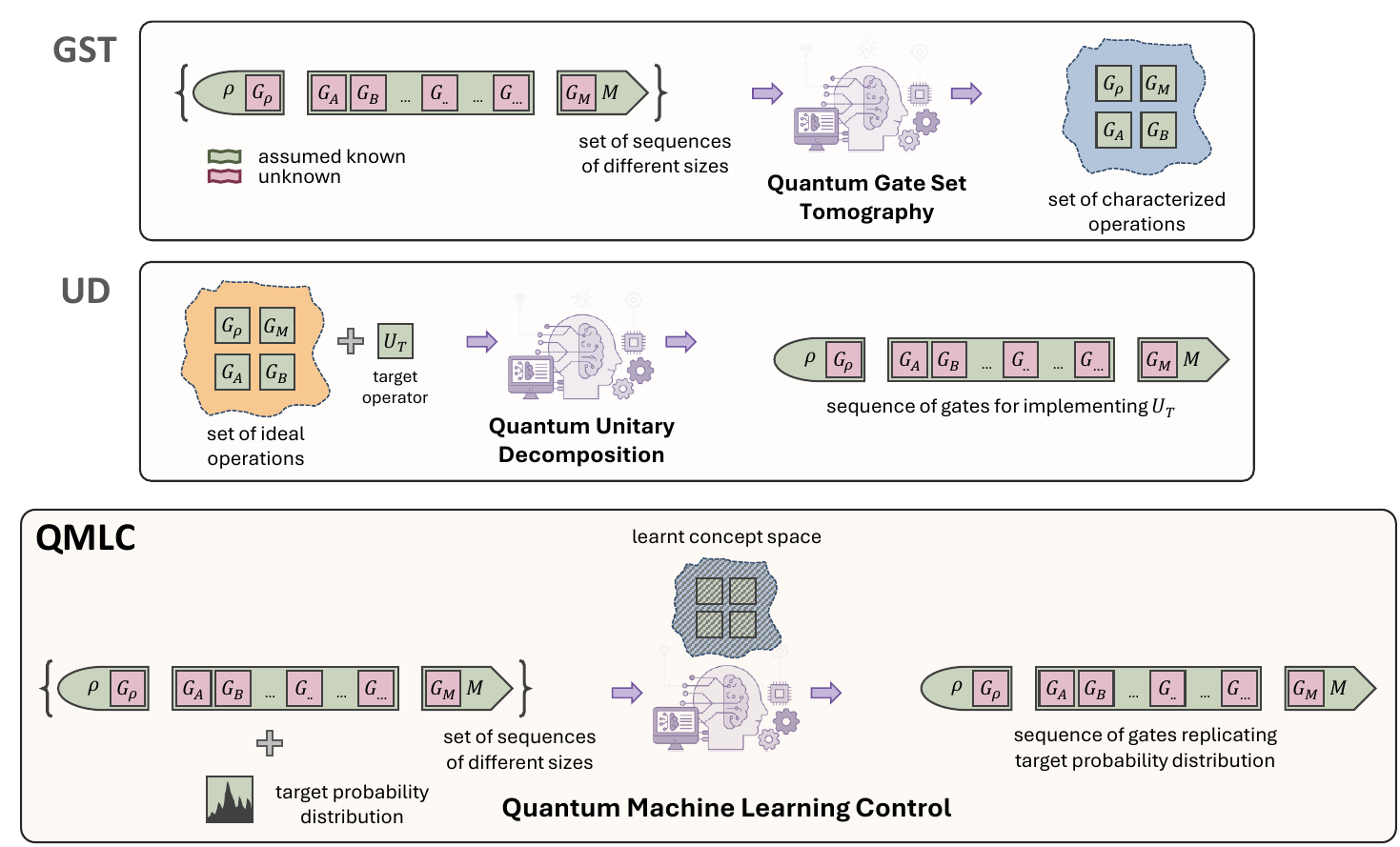}
    \caption{Workflow depicting the two traditional pipelines of characterizing native quantum gates via gate set tomography (GST) and unitary decomposition (UD) algorithm, and the proposed quantum machine learning control (QMLC) method of directly learning a generative concept space from GST data, enabling conditional synthesis of quantum circuits that produce a desired output distribution.}
    \label{fig:workflow}
\end{figure}

In this work, we explore the opportunity~\cite{hardy2020quantum} of bridging hardware characterization data and gate-based compilation in quantum computation.
To this end, we propose a generative framework that integrates gate-level tomography data directly into the quantum circuit synthesis process. 
The proposed workflow of our QMLC framework is shown in Figure~\ref{fig:workflow}.
Instead of following the traditional workflow of first using GST to characterize a tomographically complete set of gates and measurement operators (see GST block) and then performing unitary decomposition over those operators (UD block), our approach learns and maintains a latent concept space (QMLC block, short for Quantum Machine Learning Control). 
This is achieved by first tokenizing and embedding GST germ circuits, i.e., short sequences designed to amplify gate errors~\cite{greenbaum2015introduction, nielsen2020probing, nielsen2021gate}.
The training set for these circuits is organized according to a schedule inspired by curriculum learning, where each subset introduces longer, more information-rich sequences while maintaining the variability in the output statistics within the subset. 
These training sequences are processed by a set-vision transformer (Set-ViT)~\cite{lee2019set}, equipped with permutation-invariant pooling, which outputs a fixed number of k-seed vectors that capture the underlying structure of the hardware behavior, as inferred from the GST data.
The concept space is refined and sampled using a context-denoising diffusion model.
At inference time, the user provides a conditional prompt in the form of a target measurement probability distribution in the target basis, along with side information on the circuit's length bound and the subset of gates to use.
A denoising diffusion model generates a candidate circuit, which is passed through a conditional diffusion model~\cite{ding2024ccdm} trained to denoise circuit outputs in the learned distributional space.
This candidate circuit is the output of our model, which implements the target probability distribution for the specific noisy quantum hardware.

This integration of system identification, hardware-aware representation learning, generative modeling, and denoising opens the possibilities to a range of applications, from context-driven quantum algorithm adaptation to robust control pulse generation directly matched to calibrated hardware. 
Our proposed quantum machine learning control (QMLC) method also lays the groundwork for continual learning and real-time circuit re-compilation in the presence of hardware drift. 

The remainder of this article is structured as follows: Section~\ref{sec:background} provides the motivation of the proposal and the required background of the quantum technique and the ML-based models. Sections~\ref{sec:preliminaries} establish the foundational GST dataset format and our overarching research framework. Building on this, Section~\ref{sec:datasetenc} details the dataset encoding strategy, utilizing a Set-ViT to derive the latent contexts essential for the generative process. To ensure training stability and high-fidelity synthesis, Section~\ref{sec:curriculum} outlines our specialized data grouping and curriculum learning techniques. The core of our contribution is presented in Sections~\ref{sec:hierarchical} and~\ref{sec:diffusion}, where we formally define the hierarchical diffusion framework and its variational objectives. Finally, Section~\ref{sec:applications} paths towards practical utility of our QMLC framework across key applications, followed by concluding remarks in Section~\ref{sec:conclusion}.

\section{Background and Motivation}
\label{sec:background}

Quantum computation is typically expressed in the circuit model.
An algorithm is decomposed into a sequence of elementary quantum gates drawn from some basis, which are mapped to the native gate operations that can be controlled on the hardware, then scheduled and executed. 
The basis typically allows universal computation~\cite{barenco1995elementary} with local (e.g., single- and two-qubit gates) operations, and can be either discrete or continuous (e.g., single-qubit parametric rotations about a fixed axis).
The traditional approach to circuit decomposition (also called synthesis or transpilation) involves mathematically decomposing a target n-qubit unitary into a product of gates using standard techniques like quantum Shannon decomposition~\cite{krol2022efficient, krol2024beyond}, QR or cosine-sine decompositions, Cartan decomposition, Solovay-Kitaev approximations~\cite{dawson2005solovay}, or Ross-Sellinger approximations~\cite{ross2016optimal,weiden2024high}. 
These methods provide correctness (exact or approximate) at the unitary level and are relatively well understood, with performance metrics~\cite{sarkar2026yaqq} centered on effective gate count, circuit depth, and fidelity under ideal unitary operations.

However, real hardware has many non-idealities, for example, finite coherence time, restricted connectivity, noise (both coherent and incoherent), limited control over pulse shapes, cross-talk, calibration drift, and more~\cite{eisert2025mind}. 
These discrepancies mean that a circuit that is optimal under the abstract gate-unitary model does not perpetuate the fidelity in practice~\cite{waintal2024quantum}.
To bridge this gap, there are essentially two complementary directions.
Pulse-level quantum optimal control operates at a lower level of abstraction and uses microwave or laser waveforms to directly drive the qubit Hamiltonian. 
Methods in this space include optimal control algorithms (such as GRAPE~\cite{khaneja2005optimal}, EO-GRAPE~\cite{fauquenot2025open}, and GEOPE~\cite{lewis2025quantum}), firmware that combines hardware calibration and analog modulation, and approaches that aggregate circuits or gate sequences and optimize pulses in tandem~\cite{campbell2023superstaq}. 
Pulse-level control can reduce execution latency, minimize errors caused by unnecessary gate decompositions, and sometimes improve resilience against noise. 
Recent works~\cite{ball2021software, smith2022programming} in this direction propose compiler frameworks that break the standard Instruction Set Architecture (ISA) and optimize across gate aggregates to achieve speedups by exploiting pulse-level control. 
In the second approach, machine-learning-based decomposition has received considerable attention.
As quantum devices and experiments scale, using mathematically exact synthesis becomes increasingly expensive, both in terms of quantum resources and compilation tractability.
Moreover, it becomes difficult to incorporate noise and hardware constraints. 
ML-based tools are increasingly being explored to help with decomposition, e.g., (i) models that propose ans\"atze~\cite{weiden2023improving,younis2021qfast} that are close to the target unitary to reduce the search effort, (ii) reinforcement learning or other ML methods to choose instruction order, map qubits, optimize small circuit windows, or decide which decompositions to use and when to apply approximations~\cite{rietsch2024unitary,paradis2024synthetiq,belli2024scalable}, and (iii) unitary neural networks~\cite{zomorodi2024optimal} that try to directly capture input-output behavior via weights that are constrained to be unitary, are some promising directions.
While the two complementary approaches of pulse-level control and ML-based decomposition are interesting directions, their fusion into a unified layer is largely unexplored.
In particular, ref~\cite{kakkar2025no} tackles hardware-aware compilation by accepting the device's inherent ``imperfect" physics. Rather than iteratively fine-tuning pulses to suppress coherent errors or match ideal gate definitions, they explicitly characterize the native dynamics of uncalibrated gates and adapt the circuit synthesis routine to construct operations directly using these raw primitives. While effective for coherent errors, these methods typically rely on explicit Hamiltonian characterization and classical optimization, in contrast to our proposed data-driven, generative approach that implicitly learns the full noise model via GST and diffusion.
In what follows, we develop a QMLC framework that integrates the approaches from noise characterization to gate sequence compilation, and discuss its advantages.

An often necessary gateway for both decomposition and pulse control is characterizing the hardware via Gate Set Tomography (GST) or calibration processes. 
ML has also begun to be used here, for example, to model drift, detect coherent errors, predict miscalibrations, and more. 
This helps inform both traditional and ML- or pulse-based decomposition workflows. 
In ref~\cite{ho2025ai}, the authors apply standard supervised regression to GST style characterization data, with the primary aim of interpolating device noise within a predefined parameter range. Their method is intentionally lightweight and GST-inspired rather than a full, self-consistent gate set tomography procedure. In practice, they prepare tensor-product basis states, optionally apply a native gate, and record Pauli expectation values to form two matrices: \(g\) (no gate) and \(U\) (with the gate). These expectation matrices, rather than maximum likelihood reconstructions with germ sequences, serve directly as the features for learning. The gate pool is restricted to a curated, discrete set of native single qubit \(\mathrm{PRx}\) angles and \(\mathrm{CZ}\), chosen so that different noise mechanisms imprint distinguishable signatures on the resulting \(g\) and \(U\).

To make the characterization simulator ready, the authors posit a small heuristic channel family: per-qubit channels composing depolarizing, amplitude damping, dephasing, and readout errors; and, for each qubit pair, an additional two-qubit depolarizing channel composed with the single-qubit channels. 

These channels are combined after the gates in classical simulation, yielding a compact, composable device noise emulator.

Learning is performed as plain supervised regression by training two shallow multilayer perceptrons (MLPs) to learn the inverse map from \(g\) and \(U\) to the noise parameters. Inputs are the GST style expectation matrices; outputs are the corresponding channel parameters, bounded to maintain valid ranges. The training corpus is synthetic; it simulates the curated gate set by performing a multidimensional grid sweep over the noise parameters to generate $g$ and $U$, and uses the associated parameter values as labels. After training, they execute the same GST-style routine on hardware to obtain empirical \(g\) and \(U\), which are then fed into the regressors to estimate the device-specific parameters. 

Their design emphasizes interpolation over the grid, as extrapolation beyond the trained domain is not the goal and is expected to be unreliable. Meanwhile, the learned noise family is fundamentally constrained by the expressivity of the emulator's channel ansatz, because effects not encompassed by the assumed composition (e.g., coherent, non-Markovian, crosstalk-mediated errors, leakage) are effectively projected away. As a result, the practical usefulness of the fitted model is bounded by the fidelity of the underlying simulation assumptions. Any mismatch between hardware reality and the simulator induces systematic bias in the inferred parameters.

The framework employs a shortcut relative to full GST, omitting the germ sequences necessary for self-consistency. It is crucial to define this precise boundary because, unlike full GST, this approach cannot isolate gate errors from SPAM bias. Formally, treating the state initialization and measurement operators as fixed rather than variable makes the method functionally resemble Process Tomography. This limits the achievable reconstruction accuracy to the noise floor of the SPAM operations.

By contrast, Ref.~\cite{yu2025transformer} adheres to the standard GST setup and introduces \textsc{ML4GST}: a transformer-based (Vision Transformer-style) regressor that predicts gate- and device-level noise parameters from GST datasets. The network's outputs parameterize a simulated noise model, and training optimizes these parameters to match measured GST data, thus relying on the emulator to define the mapping.

While this preserves the formal GST framework, the approach remains constrained by the expressivity and fidelity of the assumed noise model and incurs additional computational cost from repeatedly simulating GST circuits during training. The usefulness of the learned parameters is therefore bounded by how accurately the emulator reflects hardware, and the inner-loop simulation cost scales linearly with the cumulative depth of the GST sequences.

Motivated by these constraints, the QMLC framework proposed in this paper bypasses both the fixed parametric noise family and the need for in-loop circuit simulation by using a generative diffusion model that directly compiles quantum circuits from user input, aiming to internalize hardware regularities in a data-driven way while improving scalability.

Recent studies have demonstrated that diffusion models~\cite{furrutter2024quantum,chen2025uditqc,barta2025leveraging,furrutter2025synthesis} are effective quantum-circuit synthesizers.  
Most notably, \cite{furrutter2024quantum} introduces a denoising diffusion probabilistic model (DDPM) whose conditioning signal is an encoded unitary matrix.  
Millions of random circuits are simulated once, each labeled with its exact unitary, giving rise to paired training samples,
\(
(C,L),
\)
where \(C\) is a gate sequence and \(L\) is its ideal unitary representation.  
During inference, the DDPM is conditioned on the label tensor \(L\), and an optional text prompt (e.g.\ ``use gate set \([\mathrm{X},\mathrm{Y},\mathrm{CX}]\)'') embedded by a frozen CLIP encoder~\cite{radford2021learning}.

The above framework relies exclusively on noise-free simulated data.  
Because simulation yields noiseless unitaries, a fully informative label \(L\) is always available.  
Real devices, however, deviate from this ideal in two critical ways: (1) Intrinsic noise, since experimental counts are inevitably corrupted by state-preparation and measurement (SPAM) errors, decoherence, and cross-talk, so the data distribution no longer matches the simulator output, and  (2) Inaccessible labels, as obtaining a full density matrix for every candidate circuit would require quantum-state tomography, which is exponentially costly in qubit number and therefore impractical for realistic training sets.

The theoretical viability of applying such data-driven approaches to quantum systems has been established in the context of machine learning control (MLC), as demonstrated by ref~\cite{hardy2020quantum}. Motivated by this precedent, and shifting the focus from control to characterization, we conclude this section by posing our primary research question:

\emph{Can a generative model, specifically a denoising diffusion probabilistic model, learn directly from real, noisy tomographic data to capture hardware characteristics without relying on idealized density-matrix labels, while generating circuits whose statistics agree with experiment?}

In this article, we propose a potential solution towards the development of a noise-aware circuit synthesis framework that integrates a permutation-invariant set transformer encoder with a conditional diffusion decoder trained on gate-set tomography data.  
GST is information-complete, i.e., in principle, it fully characterizes the underlying quantum processor through experimentally obtained outcome frequencies.  
Instead of supplying explicit labels, we embed each unordered GST dataset into a latent code that captures its joint statistics, and train the diffusion decoder to generate new circuits whose synthetic GST outputs match the empirical distribution.

This design yields an end-to-end machine-learning pipeline that dispenses with simulator-generated supervision and respects real hardware noise by construction.

\begin{figure}[htb]
    \centering
    \includegraphics[width=1.0\linewidth,trim={0cm 1.5cm 0cm 1.8cm},clip]{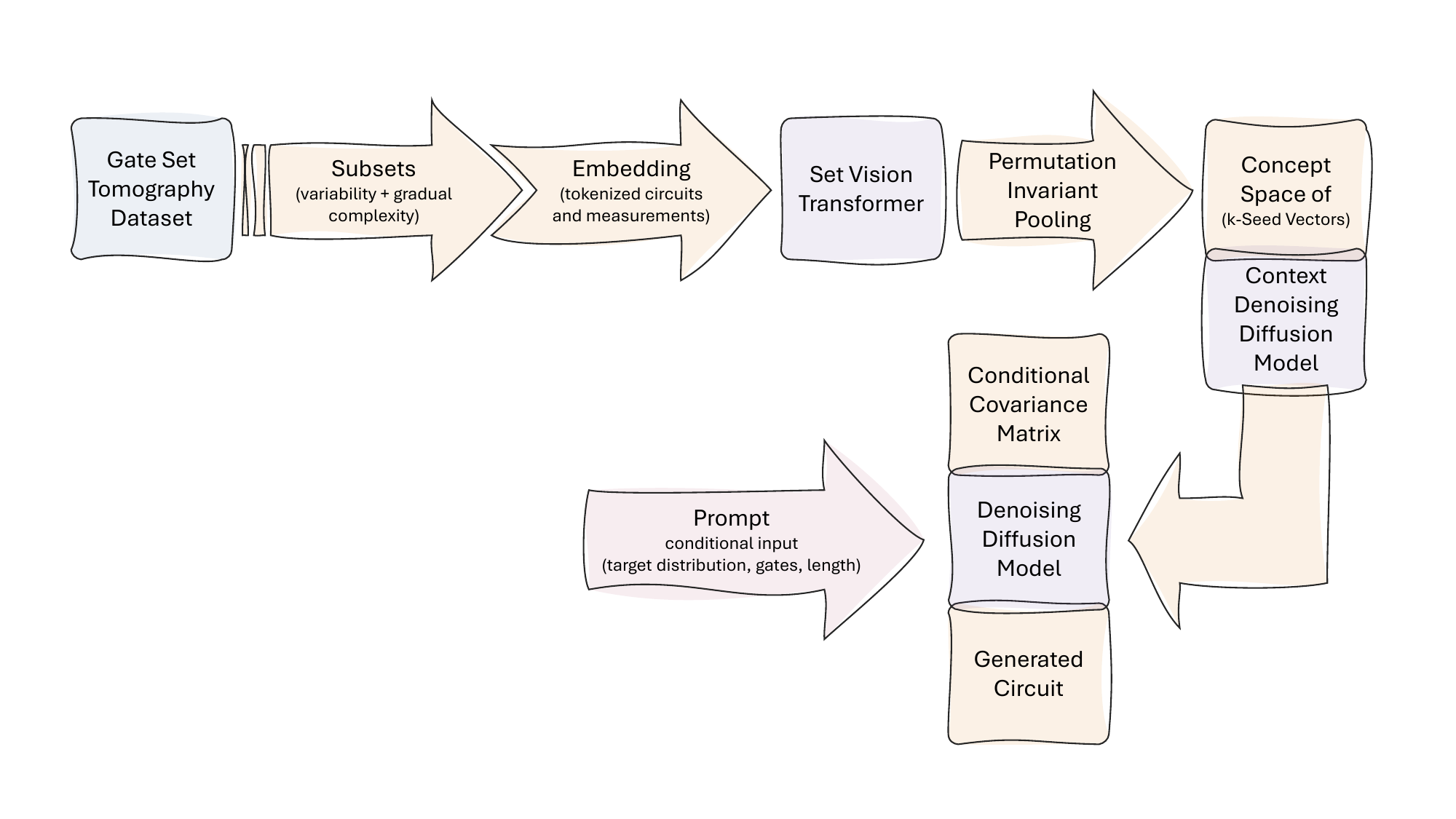}
    \caption{Overview of the proposed architecture for tomography-driven quantum circuit generation.}
    \label{fig:architectural}
\end{figure}

\section{Preliminaries}
\label{sec:preliminaries}

In the following sections, we formalize the GST-conditioned generative-compilation problem, detail the encoder-decoder architecture, present a likelihood-based training objective that enforces physical validity, and validate the framework on experimental datasets. Figure~\ref{fig:architectural} shows an overview of our architectural pipeline.

Our primary objective is to replace full-information labels $L$ with experimentally realistic Gate-Set Tomography (GST) data, all while preserving the capability for generative circuit compilation. In modern GST protocols, an experimental data point is characterized by the tuple $(C, \rho_0, M, \mathbf{p})$. Here, $C$ represents the quantum circuit, while $\rho_0$ and $M$ denote the state-preparation and measurement (SPAM) operators, respectively. The vector of outcome probabilities $\mathbf{p} \in \mathbb{R}^{d}$ of $N$ qubits, where $d = 2^N$, is obtained by executing the circuit, formally defined as
\begin{equation*}
    \mathbf{p} = (p_k)_{k=1}^d, \quad \text{where} \quad p_k = \mathrm{Tr}\bigl( M_k \, C(\rho_0) \bigr) \quad \text{with} \quad \sum_k p_k = 1.
\end{equation*}
Because quantum measurement induces wavefunction collapse, distinct physical processes can map to identical probability vectors $\mathbf{p}$. Consequently, any single pair $(C, \mathbf{p})$ carries only partial information regarding the underlying gate set. This observational constraint motivates our central inquiry: given an aggregate dataset
\begin{equation*}
    \mathcal{D}=\{(C_i, \rho_{0,i}, M_i, \mathbf{p}_i)\}_{i=1}^N,
\end{equation*}
is it possible to learn an implicit, low-dimensional latent code $\tilde{h} \approx h$ that captures the device's physical dynamics? Such a representation would allow us to decode new circuits whose predicted statistics align with experimental results.

Now consider any paired data instance
\(
(A_i,B_i) 
\) in a dataset $\mathcal{D}$, i.e. $A_i=(C_i, \rho_{0,i}, M_i)$ and $B_i=\mathbf{p}_i$
satisfying the structural equation
\[
B_i \;=\; O\;\!\bigl(A_i;\,h\bigr),
\]
with a known forward operator \(O\) (in our case, the matrix multiplication of quantum gates) and hidden parameters \(h\).
Because each pair provides only partial constraints on $h$, the inverse problem is locally underdetermined; thus, $O$ cannot be inverted via pointwise encoding alone.
We therefore propose a two-stage learning scheme:

\begin{enumerate}
  \item Dataset Encoder:  
        map the full dataset \(\{B_i\}_{i=1}^N\), together with \(\{A_i\}_{i=1}^N\) to a latent vector
        \(
        \tilde h \in \mathbb{R}^k,
        \)
        with a feature size $k$.
  \item Generative Decoder:  
        sample \(A\sim p_\theta(A\mid\tilde h)\) (implemented via a diffusion model) and accept if the simulated
        \(
        \hat B = O(A;\tilde h)
        \)
        is statistically consistent from the empirical distribution of~\(B\).
\end{enumerate}

Currently, we do not strictly enforce physical regularizations on $\tilde{h}$, such as Completely Positive Trace-Preserving (CPTP) maps for valid quantum evolution or gauge-fixing to address inherent freedom in the gate set's coordinate system. Instead, we adopt a flexible approach where extra degrees of freedom are permissible, provided the combination of $\tilde h$ and the decoder successfully reproduces the observed statistics. This flexibility simplifies the current framework, but integrating physical priors could significantly enhance both training efficiency and model interpretability in future work. Potential refinements include adding penalties for deviations from CPTP conditions or defining structured latent spaces based on physical circuit models, such as Hamiltonian generators or specific error channels, similar to how shadow tomography efficiently extracts properties from sparse quantum data.

\section{Dataset Encoding}
\label{sec:datasetenc}
A single GST circuit-measurement pair \((C,\mathbf p)\) provides only a partial and noisy observation of the underlying processor.  
Each execution probes a restricted subspace of the device's full quantum dynamics, revealing only a limited number of operator matrix elements.

Consequently, any model that conditions on a single circuit at a time risks overfitting to local idiosyncrasies (shot noise, SPAM errors) and underrepresenting global hardware regularities. For example, without access to complementary bases, a model may fail to distinguish a coherent superposition from a fully dephased mixed state, as both yield identical statistics in a single measurement setting.

To achieve informational richness, we group $n$ distinct circuits into a mini-dataset $S=\{(C_{i},\mathbf{p}_{i})\}_{i=1}^{n}$, thereby exposing the encoder to a richer, joint empirical distribution of outcomes. The resulting latent context vector acts as a sufficient statistic, summarizing shared information, such as common gate errors, correlated crosstalk, and drift, across the set. Effectively, this captures device-level physics that no single circuit can reveal in isolation. Processing such a collection requires the encoding mapping $f:(X)^n \to Z$, where $X$ is the space of individual circuit-outcome pairs and $Z$ is the continuous latent representation space, to be permutation invariant. Formally, this implies that $f(x_{\pi(1)},\dots,x_{\pi(n)}) = f(x_1,\dots,x_n)$ for every permutation $\pi \in S_n$, ensuring that the representation depends solely on the multiset of inputs rather than the order in which they are presented.

\subsection{Permutation Invariance and Non-Local Correlation}
Gate-Set Tomography (GST) produces a collection
\[
S=\{(C_i,\mathbf{p}_i)\}_{i=1}^{n}
\]
of circuit-measurement pairs.  
The physical distribution of outcomes is determined entirely by the underlying gate set and circuit arrangement. In the context of offline reconstruction, the dataset $\mathcal{S}$ is treated as a static collection of circuit-measurement pairs. Although real hardware exhibits temporal drift, rendering the physical execution order significant, we currently model the system as stationary over the collection window. Thus, our learned representation is constrained to be invariant to permutations of the input pairs. This focuses the model on the time-averaged gate behavior and global error correlations. Explicitly modeling non-stationary drift by conditioning on measurement timestamps $t$, e.g., inputs of the form $(C, p, t)$, is a promising avenue for future work, but remains outside the scope of this static reconstruction framework.

\vspace{0.25\baselineskip}
Long GST sequences encode non-local correlations that arise from coherent error accumulation and entangling operations spanning many time steps.  
Transformer self-attention naturally captures such long-range structure without inductive biases toward local connectivity (unlike convolutions)~\cite{vaswani2017attention}, making it a principled choice for quantum-physics workloads where global contextual reasoning is essential.

In summary, encoding mini-datasets as unordered sets enables the model to (i) Integrate complementary partial observations into a unified, information-rich context; (ii) Respect the physical permutation invariance of tomographic experiments; and (iii) Leverage transformer attention to model the intrinsically long-range correlations present in realistic quantum devices. These principles guide the design of our Set Vision Transformer (Set-ViT) encoder, which forms the backbone of the generative framework introduced in the following.

\subsection{Circuit and Label Embedding}

To process GST circuits and their associated measurement counts in a manner that respects set invariance, we first transform each circuit into an image-like array suitable for a Vision Transformer (ViT) encoder~\cite{dosovitskiy2020image}. Concretely, each circuit $C$ is represented as a $Q\times T$ grid of integer tokens, where $Q$ denotes the number of qubits and $T$ denotes the maximum circuit depth (number of time steps).

We begin by tokenizing the circuit, assigning a unique integer to each gate type and special symbol. For example, we define the mapping:
\[
\text{X}\;\mapsto 1,\quad
\text{Y}\;\mapsto 2,\quad
\text{CX}\;\mapsto 3,\quad
\text{Idle}\;\mapsto 4,\quad
\text{Padding}\;\mapsto 5,\;\dots
\]
This process transforms the circuit layout into an integer matrix (or grid) $G$ of size $Q \times T$, where every entry represents a specific token:
\[
G \in \{1, \dots, K\}^{Q \times T},
\]
with $K$ representing the size of the token vocabulary.

Next, we project these discrete tokens into a continuous space using an embedding function $E_{\mathrm{gate}}: \{1, \dots, K\} \to \mathbb{R}^{d_{\mathrm{gate}}}$. We strictly enforce orthogonality, such that for distinct tokens $i \neq j$, $\langle E_{\mathrm{gate}}(i), E_{\mathrm{gate}}(j) \rangle = 0$ and $\|E_{\mathrm{gate}}(i)\| = 1$. Applying this embedding element-wise to the integer grid converts $G$ into a real-valued circuit tensor $\mathcal{G}$:
\[
\mathcal{G}_{q,t} = E_{\mathrm{gate}}(G_{q,t}) \quad \implies \quad \mathcal{G} \in \mathbb{R}^{Q \times T \times d_{\mathrm{gate}}}.
\]
This resulting 3D array encodes the qubit index, time step, and gate type along its three axes.

Simultaneously, the measurement counts $\mathbf m \in \mathbb{R}^{\,d_m}$ associated with each circuit execution are normalized by their $\ell_{1}$‐norm to form a vector
\[
\mathbf y \;=\; \frac{\mathbf m}{\|\mathbf m\|_{1}}
\;\in\;\mathbb{R}^{\,d_m}.
\]

While the normalized vector $y$ faithfully represents the empirical measurement distribution, directly feeding it into a neural network poses a subtle but critical challenge related to the physics of our quantum system. In the ideal noiseless case, the target measurement probability $y$ for Clifford circuits is highly structured, distributing uniformly over affine subspaces of the Boolean hypercube. Consequently, the vector is dominated by a few ``major'' terms (e.g., $0.5$ or $0.25$) and zeros. However, the critical physical information we seek to detect, such as SPAM errors, coherent miscalibrations, and cross-talk, resides entirely within the minor perturbation terms, $y_i = y_{i,\text{ideal}} \pm \delta_i$.

Standard Multi-Layer Perceptrons (MLPs) are known to exhibit a strong \textit{spectral bias}~\cite{Rahaman2018OnTS}, meaning they preferentially learn low-frequency, high-amplitude functions. If $y$ is ingested directly, the network will easily capture the major structural distribution but critically smooth over or ignore the high-frequency $\delta_i$ deviations, thereby losing the subtle hardware noise signatures. 

To resolve this, we propose applying a pre-conditioning mathematical transformation $\mathcal{T}: \mathbb{R}^{d_m} \to \mathbb{R}^{d_{\text{emb}}}$ to the probability vector, $\tilde{y} = \mathcal{T}(y)$, designed specifically to amplify these physically significant minor terms. Within our framework, this transformation can be instantiated using one of three theoretically motivated encoding strategies:

\begin{enumerate}
    \item \textbf{Boundary-Amplified Logit Embedding:} In quantum physics, an error $\delta$ on an outcome that should theoretically be strictly forbidden (probability 0) carries disproportionately high Shannon information regarding leakage or bit-flip mechanisms. To aggressively stretch the vector space near zero and highlight these specific errors, we apply a clamped logit transformation element-wise:
    \begin{equation*}
        \tilde{y}_i = \log\left(\frac{y_i + \epsilon}{1 - y_i + \epsilon}\right),
    \end{equation*}
    where $\epsilon \approx 10^{-6}$ acts as a numerical stabilizer. This mapping converts a minuscule physical perturbation (e.g., $\delta = 10^{-3}$) into a significantly shifted continuous embedding value, forcing the subsequent network layers to attend heavily to small leakage terms.

    \item \textbf{Walsh-Hadamard Transform (WHT):} Alternatively, we can map the probability distribution from the computational basis into the space of Pauli expectation values. The mathematically rigorous equivalent of a Fourier Transform on the Boolean hypercube $\Omega_Q$ is the Walsh-Hadamard Transform~\cite{ODonnell2014AnalysisOB, WalshACS}:
    \begin{equation*}
        \tilde{y} = H^{\otimes Q} y,
    \end{equation*}
    where $H$ is the standard $2\times 2$ Hadamard matrix. By stabilizer theory, an ideal Clifford output yields exact expectation values of $\{0, \pm 1\}$. Under real hardware noise, these values become $\pm (1 - \delta_1)$ and $\delta_2$. The WHT effectively isolates the physical error terms into distinct parity dimensions, naturally decoupling the major structural signals from the minor physical perturbations.

    \item \textbf{High-Frequency Sinusoidal Representation:} To counteract the MLP's spectral bias directly, we can adapt the concept of positional encoding used in standard Transformers to our continuous probability features. We project each probability element $y_i$ into a high-dimensional, high-frequency Fourier feature space~\cite{Tancik2020FourierFL}:
    \begin{equation*}
        \tilde{y}_i = \bigoplus_{j=0}^{L-1} \left[ \sin(2^j \pi y_i), \cos(2^j \pi y_i) \right],
    \end{equation*}
    where $L$ controls the maximum frequency band and $\oplus$ denotes concatenation. Because the frequency $2^j$ acts as a massive exponential multiplier inside the trigonometric argument, even an infinitesimal shift $\delta_i$ results in a detectable, orthogonal variation in the higher-dimensional embedding space, guaranteeing that the network detects the subtle error mechanics.
\end{enumerate}

Following the methodology of the original CCDM paper~\cite{ding2024ccdm} to enable label consistency, we then feed the transformed distribution $\tilde{y}$ (rather than the raw $y$) into a deep multilayer perceptron $f_\theta$ of at least five hidden layers, obtaining a new vector $h_y^l$ of embedding dimension $d_{\text{circuit}}$:
\begin{equation*}
    h_y^l = f_\theta(\tilde{y}) \in \mathbb{R}^{d_{\text{circuit}}}.
\end{equation*}

 This long label embedding $\bm h_{y}^{l}$ has the same dimension as the flattened clean (circuit) image $\bm x_0$ in the diffusion model, and is later used for constructing the covariance matrix during diffusion. Likewise, we can obtain the short-label embedding $\bm h_{y}^{s}$, which serves as the typical input to the set-ViT encoder and as the conditional injection for the diffusion model, by following similar steps. For completeness, we briefly outline the procedure for obtaining the long-label embedding $\bm h_{y}^{l}$ in the CCDM framework, introducing the concept of label consistency that stabilizes training.

We introduce three neural subnetworks for label consistency:
\[
T_{1}:\;\bm x \;\mapsto\;\bm h_{y}^{l},\qquad
T_{2}:\;\bm h_{y}^{l} \;\mapsto\;\widehat{\mathbf y},\qquad
T_{3}:\;\widehat{\mathbf y}\;\mapsto\;\widehat{\bm h}_{y}^{l}.
\]
Here, $T_{1}$ and $T_{2}$ are trained jointly to predict normalized probability $\widehat{\mathbf y}$ from the circuit image $\bm{x}$, and once converged, they are frozen while $T_{3}$ is trained via the CCDM consistency loss.  Specifically, for a mini-batch of size $m$, we add standard multidimensional Gaussian noise
\[
\bm\gamma \;\sim\; \mathcal{N}\bigl(\mathbf 0,\,\sigma^{2}\bm I\bigr)
\]
to each true label $\mathbf y_{i}$ and minimize
\[
\frac{1}{m}\sum_{i=1}^{m}
\mathbb{E}_{\bm\gamma}\Bigl[\,
  \bigl\|
    T_{2}\bigl(T_{3}(\mathbf y_{i} + \bm\gamma)\bigr)
    \;-\;(\mathbf y_{i} + \bm\gamma)
  \bigr\|_{2}^{2}
\Bigr].
\]

This consistency loss ensures that $T_{3}$ learns a semantically meaningful inversion of $T_{2}$ rather than producing arbitrary latent features.

\subsection{Set Vision Transformer Encoder}

After obtaining the 3D circuit array $\mathcal{G}$ and the short label embedding $\bm h_{y}^{s}$, we feed them into a Vision Transformer (ViT) encoder.  The steps are as follows:

\begin{enumerate}
  \item \textbf{Patch Embedding.}  
  The circuit array of shape $(Q,T,d_{\mathrm{gate}})$ is partitioned into non-overlapping patches of size $2\times 2$ (two qubits by two time steps).  For instance, if $Q=2$ and $T=20$, this yields $\frac{Q}{2}\times\frac{T}{2}=10$ patches along the time dimension.  Each $2\times 2$ patch is a tensor in $\mathbb{R}^{2\times 2\times d_{\mathrm{gate}}}$, which is flattened into a vector in $\mathbb{R}^{4\,d_{\mathrm{gate}}}$.  We then apply a learnable linear projection
  \[
    \mathrm{PatchEmbed}:\;\mathbb{R}^{\,4\,d_{\mathrm{gate}}}
    \;\longrightarrow\;\mathbb{R}^{\,d_{\mathrm{model}}},
  \]
  producing a sequence of $10$ token embeddings 
  \(\{\bm e_{k}\}_{k=1}^{10}\subset\mathbb{R}^{\,d_{\mathrm{model}}}.\)

  \item \textbf{Tokenization.}  Following ref.~\cite{dosovitskiy2020image}, we introduce a single learnable token \texttt{[LRN]} in the manner of the original Vision Transformer.  Specifically, we prepend the learnable token 
\[
\texttt{[LRN]} \;\in\; \mathbb{R}^{\,d_{\mathrm{model}}}
\]
at the beginning of the tokenized sequence.  Immediately thereafter, we place the label token 
\[
\texttt{[LBL]}\;\equiv\bm h_{y}^{s}\ ; \;\in\; \mathbb{R}^{\,d_{\mathrm{model}}},
\]
which is not learned directly but is produced by the pre-trained network \(T_{3}\).  Thus, our token sequence becomes
\[
  \bigl[\;\texttt{[LRN]};\;\texttt{[LBL]};\;\bm e_{1};\dots;\bm e_{10}\bigr]
  \;\in\;\mathbb{R}^{\,(T+2)\times d_{\mathrm{model}}},
\]
where $\{\bm e_{k}\}$ are the $2\times 2$ patch embeddings.  
The learnable token \texttt{[LRN]} serves as a global summary vector exactly as in the original ViT formulation, while \texttt{[LBL]} carries the encoded measurement information from \(T_{3}\).

  \item \textbf{Positional Encoding.}  We add a learnable positional embedding $E_{\mathrm{pos}}[k]\in\mathbb{R}^{\,d_{\mathrm{model}}}$ to each token, for indices $k=0,\dots,T+1$, where $k=0$ corresponds to $\texttt{[LRN]}$, $k=1$ to $\texttt{[LBL]}$, and $k=2,\dots,T+1$ to patches $1,\dots,T$, respectively.  This ensures the transformer sees each token's location within the circuit.

  \item \textbf{Transformer Encoder Blocks.}  
We pass the entire $(T+2)\times d_{\mathrm{model}}$ token matrix $\bm X\in\mathbb{R}^{(T+2)\times d_{\mathrm{model}}}$ through $L_{\mathrm{ViT}}$ number of identical Transformer encoder blocks.  Each block applies:
\[
  \begin{aligned}
    &\bm X' = \mathrm{LayerNorm}\bigl(\bm X + \mathrm{MultiHeadAttn}(\bm X,\bm X,\bm X)\bigr),\\
    &\bm X'' = \mathrm{LayerNorm}\bigl(\bm X' + \mathrm{rFF}(\bm X')\bigr)
  \end{aligned}
\]
  Here, MultiHeadAttn refers to a standard multi-head attention layer, and rFF denotes a row-wise feedforward layer as defined in the Set Transformer paper\cite{lee2019set}, meaning it processes each token embedding (each row of $\bm X'$) independently through the same two-layer MLP, followed by layer normalization.  After $L_{\mathrm{ViT}}$ such layers, we take the final hidden state corresponding to $\texttt{[LRN]}$, denoted
\[
  \bm h_{LRN} \;=\; [\texttt{[LRN]}]_{\text{output}}
  \;\in\;\mathbb{R}^{\,d_{\mathrm{model}}},
\]
which serves as the high-level latent feature for both the circuit and the label. The final hidden state of $\texttt{[LBL]}$ can be stored and used for downstream tasks like diffusion.

\end{enumerate}

Because the self-attention layers operate on the entire token row, the $\texttt{[LRN]}$, $\texttt{[LBL]}$, and patch tokens can attend to one another fully.  We exploit the learned $\texttt{[LRN]}$ token as a summary of the entire circuit-label input.  After encoding a mini-set of $n$ circuit-label pairs, we collect the $\texttt{[LRN]}$ outputs into a matrix 
$\bm H_{LRN} \in \mathbb{R}^{\,n\times d_{\mathrm{model}}}$.

To maintain permutation invariance while circumventing the $\mathcal{O}(n^2)$ computational bottleneck of standard self-attention, we employ Induced Set Attention Blocks (ISAB)~\cite{lee2019set}. By mediating interactions through a set of $m$ learnable inducing points, ISAB reduces the complexity to $\mathcal{O}(nm)$, enabling the efficient processing of large circuit sets. Specifically, each $\bm H_{LRN}$ is processed by two stacked ISAB layers with $m=32$. This value was selected based on empirical evidence demonstrating that set representations saturate rapidly with respect to $m$. A value of $m=32$ provides a sufficient number of 'prototypes' to capture the global data topology while minimizing the computational overhead of the linear attention mechanism.  We then apply Pooling by Multi-head Attention (PMA) with $k$ learnable seed vectors.  For example, if the GST dataset has four unique underlying gates, we set $k=4$.  The PMA output is a set of $k$ context vectors, each in $\mathbb{R}^{\,d_{\mathrm{model}}}$.  We denote the resulting context matrix by 
$\ \bm H_{context} \in \mathbb{R}^{\,k\times d_{\mathrm{model}}}$.

\section{Data Grouping and Curriculum Learning}
\label{sec:curriculum}

\begin{figure}[t]
\centering
\begin{tikzpicture}[
    font=\small,
    box/.style={draw, rectangle, rounded corners, minimum width=12.0cm, minimum height=1.25cm, align=center, fill=gray!15},
    band/.style={draw, rectangle, rounded corners, dashed, minimum width=3.8cm, minimum height=2.6cm, align=center},
    setnode/.style={draw, circle, minimum size=0.9cm, align=center, fill=blue!15},
    stage/.style={draw, rectangle, rounded corners, minimum width=4.4cm, minimum height=1.05cm, align=center, fill=green!15},
    arr/.style={-Latex, very thick}
]
\node[box] (raw) at (0,0) {Raw Data Instances};

\coordinate (xL) at (-5.0,0);
\coordinate (xC) at ( 0.0,0);
\coordinate (xR) at ( 5.0,0);

\draw[arr] ([yshift=-0.1cm]raw.south -| xL) -- +(0,-1.0)
      node[midway, xshift=-1.3cm] {\scriptsize short circuits};
\draw[arr] ([yshift=-0.1cm]raw.south -| xC) -- +(0,-1.0)
      node[midway, xshift= -1.3cm] {\scriptsize medium circuits};
\draw[arr] ([yshift=-0.1cm]raw.south -| xR) -- +(0,-1.0)
      node[midway, xshift= -1.3cm] {\scriptsize long circuits};

\node[band] (bandS) at (-5.0,-2.7) {};
\node[band] (bandM) at ( 0.0,-2.7) {};
\node[band] (bandL) at ( 5.0,-2.7) {};


\node[setnode] (s1) at ($(bandS.center)+(-0.9,0.2)$) {$S_{1}$};
\node[setnode] (s2) at ($(bandS.center)+(-0.2,0.0)$) {$S_{2}$};
\node[setnode] (s3) at ($(bandS.center)+(+0.6,0.25)$) {$S_{3}$};
\node at ($(bandS.center)+(1.2,-0.55)$) {$\cdots$};

\node[setnode] (m1) at ($(bandM.center)+(-1.0,0.0)$) {$S_{k}$};
\node[setnode] (m2) at ($(bandM.center)+(-0.2,0.3)$) {$S_{k+1}$};
\node[setnode] (m3) at ($(bandM.center)+(+0.7,0.1)$) {$S_{k+2}$};
\node at ($(bandM.center)+(1.2,-0.55)$) {$\cdots$};

\node[setnode] (l1) at ($(bandL.center)+(-0.9,0.15)$) {$S_{r}$};
\node[setnode] (l2) at ($(bandL.center)+(-0.1,0.0)$) {$S_{r+1}$};
\node[setnode] (l3) at ($(bandL.center)+(+0.7,0.25)$) {$S_{r+2}$};
\node at ($(bandL.center)+(1.2,-0.55)$) {$\cdots$};

\node at (0.0,-3.5) {\scriptsize Overlap among sets is allowed};

\node[stage] (st1) at (-5.0,-6.0) {Stage 1: Short / Simple};
\node[stage] (st2) at ( 0.0,-6.0) {Stage 2: Medium};
\node[stage] (st3) at ( 5.0,-6.0) {Stage 3: Long / Complex};

\draw[arr] (bandS.south) -- (st1.north);
\draw[arr] (bandM.south) -- (st2.north);
\draw[arr] (bandL.south) -- (st3.north);

\draw[arr] (st1.east) -- (st2.west);
\draw[arr] (st2.east) -- (st3.west);

\end{tikzpicture}

    \caption{Schematic illustration of the data grouping strategy. Raw data instances are partitioned into sets such that instances within a set are maximally distinct yet share a narrow range of circuit lengths. Overlap across sets is allowed. The sets are then ordered by circuit length to enable curriculum learning from simple to complex circuits.}
    \label{fig:data_grouping}
\end{figure}
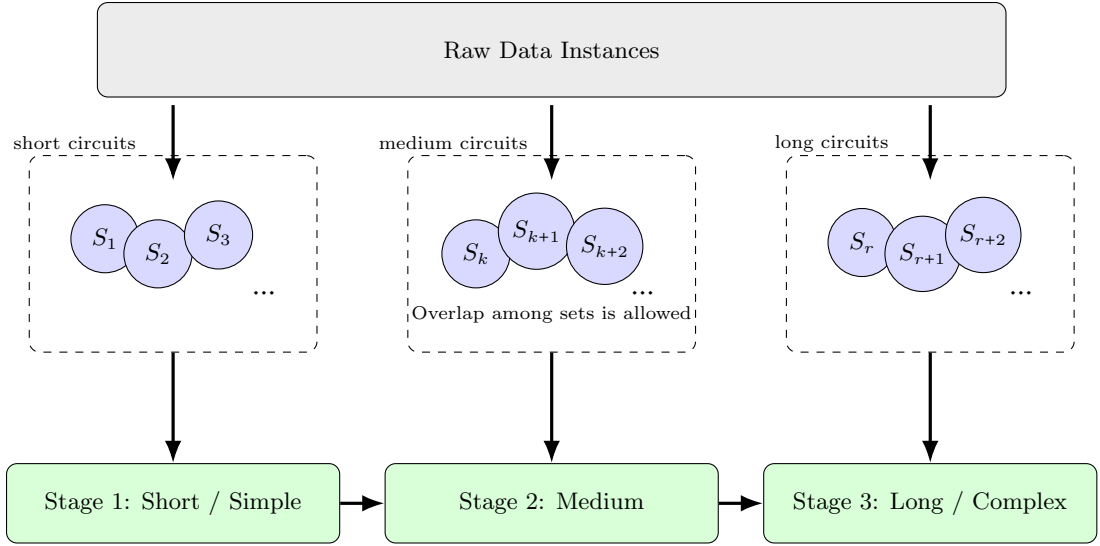

Since the proposed machine learning model is based on ``sets'' of input data (with the encoder implemented as a set transformer), it is essential to group the raw data into appropriate sets. These sets must satisfy several criteria to (i) minimize ambiguity in the downstream generative task and (ii) enhance training efficiency and quality through curriculum learning. The grouping procedure is illustrated in Figure~\ref{fig:data_grouping} and follows the principles outlined below:

\begin{enumerate}
    \item \textbf{Minimizing ambiguity.} As discussed earlier, when using GST data as the training dataset, the conditional prompt (e.g., probability distribution, gate type, circuit length) used to guide the generative model for quantum circuit synthesis is often ambiguous and degenerate (not to be confused with quantum degeneracy). That is, the same conditional prompt can correspond to multiple valid quantum circuits. To reduce such ambiguity, we group data into sets such that each data instance within a set is maximally distinct. 

    For an $n$-qubit register, projective measurement in the computational basis yields one of $2^n$ outcomes, i.e., strings in $\{0,1\}^n$. We denote the outcome set by
    \[
    \Omega_n = \{0,1\}^n, \qquad |\Omega_n| = 2^n.
    \]
    When restricting to Clifford circuits (e.g., generated by $\{X_{\pi/2}, Y_{\pi/2}, \mathrm{CNOT}\}$), the measurement distributions are not arbitrary. By stabilizer theory, any Clifford circuit maps $\lvert 0 \rangle^{\otimes n}$ to a stabilizer state whose computational ($Z$-)basis outcomes form an affine subspace of $\Omega_n$ and are distributed uniformly on that subspace.
    In this case, every distribution corresponds to a uniform distribution over an affine subspace of $\Omega_n$.
    If the affine subspace has dimension $k$, its support size is $2^k$, and each outcome occurs with probability $2^{-k}$.
    
    The number of distinct affine subspaces of dimension $k$ is
    \[
    N(n,k) \;=\; 2^{\,n-k} 
    \binom{n}{k}_2
    \]

    where $\binom{n}{k}_2$ denotes the Gaussian binomial coefficient
    \[
    \binom{n}{k}_2 = \prod_{i=0}^{k-1} \frac{1 - 2^{\,n-i}}{1 - 2^{\,i+1}}.
    \]

    Hence, the total number of distinct computational-basis measurement distributions achievable with Clifford circuits is
    \[
    T(n) \;=\; \sum_{k=0}^n 2^{\,n-k}\binom{n}{k}_2
    \]
    
     in an ideal noiseless scenario, assuming the circuit depth is sufficiently large. Accordingly, we ensure that each selected data instance within a set is distinct with respect to its probability distribution. This distinction can be further extended to include gate types and circuit lengths, thereby compounding the diversity within each set.

    \item \textbf{Circuit length consistency.} Each data instance within a set must fall within a narrow range of circuit lengths, i.e.,
    \[
    l_{\min} \leq \text{circuit length} \leq l_{\max}, \quad \text{with } l_{\max} - l_{\min} \leq \tau,
    \]
    where $\tau$ denotes a threshold parameter controlling the allowable variation in length.

    \item \textbf{Overlap across sets.} Data instances may appear in multiple sets. That is, partial overlap between sets is permitted and, in practice, beneficial for maintaining training diversity.

    \item \textbf{Curriculum learning.} The model is trained in stages according to increasing circuit length. Training proceeds from shorter and simpler circuits to longer and more complex circuits, following the principles of curriculum learning~\cite{bengio2009curriculum}\cite{soviany2022curriculum}. This staged approach improves convergence stability and helps the model progressively adapt to increasingly challenging synthesis tasks.
\end{enumerate}
Ultimately, this grouping strategy organizes the data to prevent redundancy while stabilizing the training process. We feed the network diverse sets of uniform-length circuits to ensure the learned representation captures the actual structural properties of the Clifford group, instead of simply memorizing trivial patterns or getting lost in ambiguous data.

\section{Hierarchical Diffusion Framework}
\label{sec:hierarchical}

To effectively model the complex, high-dimensional distribution of the dataset, we propose a hierarchical diffusion framework. This approach is motivated by the observation that our data is not independent and identically distributed (i.i.d.) at the global level, instead it consists of exchangeable ``mini-sets'' governed by shared underlying physical parameters.

\subsection{Theoretical Motivation}
Directly modeling the joint probability distribution $p(\bm{X})$ over the raw data space is computationally intractable and prone to mode collapse due to the underlying physics' multimodality. Instead, we decompose the generation process into two distinct semantic levels: a global \textit{context} manifold and a local \textit{state} manifold.

Let a mini-set be denoted as $\mathcal{X} = \{\bm{x}_1, \dots, \bm{x}_N\}$, where the elements are permutation invariant. We posit that the generation of $\mathcal{X}$ depends on a latent context variable $\bm{c}$ that encapsulates the global physical constraints (e.g., Hamiltonian parameters). The joint distribution factorizes as:
\begin{equation}
    p(\mathcal{X}) = \int p(\mathcal{X} | \bm{c}) p(\bm{c}) \, d\bm{c} \approx \int \left( \prod_{i=1}^N p(\bm{x}_i | \bm{c}) \right) p(\bm{c}) \, d\bm{c}
\end{equation}
This formulation necessitates a two-stage generative process: (i) Global-Context Diffusion (GCD) that models the prior distribution of physical contexts $p(\bm{c})$, and (ii) Conditional Circuit-Token Diffusion (CTD) that models the conditional likelihood $p(\bm{x} | \bm{c})$, generating detailed data structures given the physical context.

Crucially, to extract $\bm{c}$ from raw data during training, we require an encoder $E_{\text{set}}(\cdot)$ that respects the permutation invariance of the input set. A standard Convolutional Neural Network (CNN) or Multilayer Perceptron (MLP) would impose an arbitrary ordering, introducing inductive bias errors. Therefore, we employ a Set Vision Transformer (Set ViT) to serve as a permutation-invariant statistic extractor.

\subsection{Architecture}
The framework consists of three coupled components:

\paragraph{1. Set Vision Transformer (Set ViT):}
The Set ViT acts as the encoder $E_{\phi}: \mathbb{R}^{N \times d_{\text{in}}} \to \mathbb{R}^{d_{\text{ctx}}}$. Given an input mini-set $\mathcal{X}$, it aggregates information via self-attention mechanisms to produce a global context vector $\bm{z}_{\text{ctx}} = E_{\phi}(\mathcal{X})$. This vector serves as a sufficient statistic for the mini-set.

\paragraph{2. Global-Context Diffusion (GCD):}
This module operates in the latent space of the Set ViT. It is an unconditional diffusion model trained to reverse a noise process on the context vector $\bm{z}_{\text{ctx}}$. It utilizes a standard isotropic Gaussian prior $p(\bm{z}_T) = \mathcal{N}(\mathbf{0}, \bm{I})$.

\paragraph{3. Conditional Circuit-Token Diffusion (CTD):}
The CTD generates the high-dimensional circuit tokens $\bm{x}_0 \in \mathbb{R}^{L \times d_{\text{tok}}}$ conditioned on the clean context $\bm{z}_{\text{ctx}}$. Unlike standard Gaussian diffusion, the CTD incorporates a label-dependent covariance $\bm{H}_y$ into the transition dynamics to enforce measurement probability constraints. $\bm{H}_y$ is derived from a pre-trained, frozen MLP and ensures that generated samples adhere to physical validity.

\subsection{Joint Latent Diffusion Optimization}
A significant challenge in hierarchical diffusion models is end-to-end differentiability. Naively ``daisy-chaining'' the models (i.e., sampling a context $\hat{\bm{z}}$ from GCD and feeding it to CTD during training)renders backpropagation intractable, as it requires differentiating with respect to the ODE solver of the GCD sampling process.

To circumvent this, we adopt a joint training strategy inspired by Score-based Generative Modeling in Latent Space (LSGM)~\cite{vahdat2021score}. Instead of sampling from the prior during training, we use a teacher-forcing approach in which the encoder's output directly conditions the diffusion objectives. The procedure is defined as follows:

The training procedure begins with the context encoding phase, where the Set ViT processes a mini-set of real data $\mathcal{X}_{\text{real}}$ to generate a ``ground truth'' context embedding $\bm{z}_{\text{gt}} = E_{\phi}(\mathcal{X}_{\text{real}})$. This embedding serves as the anchor for two parallel diffusion objectives. On one hand, the GCD is trained to denoise $\bm{z}_{\text{gt}}$ directly, treating the latent vector itself as the data to be modeled. Simultaneously, the CTD is trained to reconstruct the original data $\bm{x} \in \mathcal{X}_{\text{real}}$ by denoising it, conditioned on this fixed ground-truth embedding $\bm{z}_{\text{gt}}$.

This configuration creates a shared information bottleneck. The gradient flow connects the Set ViT to both diffusion models simultaneously. The Set ViT is updated to produce embeddings that are both generable (low GCD loss) and informative for reconstruction (low CTD loss). Formally, the total objective is:
\begin{equation}
    \mathcal{L}_{\text{total}} = \mathcal{L}_{\text{GCD}}(\bm{z}_{\text{gt}}) + \lambda \mathcal{L}_{\text{CTD}}(\bm{x}, \bm{z}_{\text{gt}}).
\end{equation}
This ensures end-to-end differentiability without the need to backpropagate through the sampling chain.

\section{Variational Diffusion Formalism}
\label{sec:diffusion}

We adopt the continuous-time Variational Diffusion Model (VDM) framework \cite{kingma2021variational} for both the global context and the circuit tokens. This formalism allows us to jointly optimize the generative model and the noise schedule, expressed via the signal-to-noise ratio (SNR).

\subsection{Global-Context Variational Diffusion (GCD)}
\label{sec:gcd}

The GCD models the distribution of the latent context vectors $\bm{z}^{(c)} \in \mathbb{R}^{d_{\text{ctx}}}$. We define a monotonically decreasing log-signal-to-noise ratio schedule $\gamma_{\eta}^{(c)}(t): [0,1] \to \mathbb{R}$, parameterized by $\eta$. The diffusion process is governed by the noise variance $\sigma_t^{(c)2} = \operatorname{sigmoid}(\gamma_{\eta}^{(c)}(t))$ and the signal preservation coefficient $\alpha_t^{(c)2} = 1 - \sigma_t^{(c)2}$. By satisfying the condition $\alpha_t^{(c)2} + \sigma_t^{(c)2} = 1$, this coefficient ensures the forward process preserves unit variance, creating a smooth interpolation between the data distribution at $t=0$ and the standard Gaussian prior at $t=1$.

The forward diffusion kernel is given by the marginal distribution at any time $t \in [0,1]$. For a clean context $\bm{z}^{(c)}_0$, the intermediate noisy latent state $\bm{z}^{(c)}_t$ is sampled via:
\begin{equation}
    q(\bm{z}^{(c)}_t \mid \bm{z}^{(c)}_0) = \mathcal{N}(\bm{z}^{(c)}_t; \alpha_t^{(c)}\bm{z}^{(c)}_0, \sigma_t^{(c)2}\bm{I}).
\end{equation}
Equivalently, using the reparameterization trick with $\bm{\epsilon} \sim \mathcal{N}(\mathbf{0}, \bm{I})$, this can be written as the stochastic trajectory:
\begin{equation}
    \bm{z}^{(c)}_t = \alpha_t^{(c)}\bm{z}^{(c)}_0 + \sigma_t^{(c)}\bm{\epsilon}.
\end{equation}

The generative model learns to reverse this process by predicting the noise $\bm{\epsilon}$ from the noisy state $\bm{z}^{(c)}_t$. The variational lower bound on the likelihood reduces to a weighted mean squared error (MSE). As $T \to \infty$, the objective approximates:
\begin{equation}
    \mathcal{L}_{\text{GCD}} = \frac{1}{2} \mathbb{E}_{t \sim \mathcal{U}(0,1), \bm{\epsilon} \sim \mathcal{N}(\mathbf{0}, \bm{I})} \left[ \gamma_{\eta}^{(c)'}(t) \, \left\| \bm{\epsilon}_{\theta}^{(c)}(\bm{z}^{(c)}_t, t) - \bm{\epsilon} \right\|_2^2 \right].
\end{equation}
Here, $\gamma'(t)$ acts as a weighting function that emphasizes time steps where the noise level changes most rapidly. In practice, we compute the raw MSE, $\texttt{mse}^{(c)} = \| \bm{\epsilon}_{\theta}^{(c)} - \bm{\epsilon} \|_2^2$, and scale gradients by $\gamma'(t)/2$.

\subsection{Conditional Circuit-Token Variational Diffusion (CTD)}
\label{sec:ctd}

The CTD generates the token grid $\bm{x} \in \mathbb{R}^{L \times d_{\text{tok}}}$ conditioned on the label $\bm{y}$ and the clean ground-truth context $\bm{z}^{(c)}_0$. Unlike isotropic diffusion in GCD, the circuit tokens must adhere to specific physical measurement constraints encoded in the covariance matrix $\bm{H}_y$. We incorporate this into the diffusion kernel to enforce valid optimization trajectories. Given a schedule $\gamma_{\eta}(t)$, the transition kernel is anisotropic:
\begin{equation}
    q(\bm{x}_t \mid \bm{x}_0, \bm{y}) = \mathcal{N}(\bm{x}_t; \alpha_t\bm{x}_0, \sigma_t^2\bm{H}_y).
\end{equation}
Since $\bm{H}_y$ is positive definite, it introduces a directional bias to the diffusion process. The matrix scales the noise magnitude along different axes, ensuring that exploration is concentrated within the subspace of valid physical constraints. Sampling is performed by scaling standard Gaussian noise:
\begin{equation}
    \bm{x}_t = \alpha_t\bm{x}_0 + \sigma_t \bm{H}_y^{1/2}\bm{\epsilon}, \quad \text{where } \bm{\epsilon} \sim \mathcal{N}(\mathbf{0}, \bm{I}).
\end{equation}

\subsubsection{Whitened Diffusion Objective}
The standard diffusion loss must be adjusted to account for the correlated noise structure. Deriving the VDM bound for this anisotropic process leads to a Mahalanobis distance metric $\|\cdot\|_{\bm{H}_y^{-1}}^2$. To simplify computation, we ``whiten'' the error by pre-multiplying with $\bm{H}_y^{-1/2}$:
\begin{equation}
    \mathcal{L}_{\text{CTD}} = \frac{1}{2} \mathbb{E}_{\substack{t \sim \mathcal{U}(0,1) \\ \bm{\epsilon} \sim \mathcal{N}(\mathbf{0}, \bm{I})}} \left[ \gamma_{\eta}'(t) \, \left\| \bm{H}_y^{-1/2} \bm{\epsilon}_{\phi}(\bm{x}_t, t, \bm{y}, \bm{z}^{(c)}_0) - \bm{H}_y^{-1/2}\bm{\epsilon} \right\|_2^2 \right].
\end{equation}
Intuitively, this objective penalizes errors more heavily in directions where the variance $\bm{H}_y$ is small (high certainty constraints) and less in directions of high variance. Since $\bm{H}_y$ is diagonal in our implementation, this operation is efficiently computed as an element-wise division:
\begin{equation}
    \texttt{mse}_{\text{white}} = \left\| \frac{\bm{\epsilon}_{\phi} - \bm{\epsilon}}{\sqrt{\operatorname{diag}(\bm{H}_y)}} \right\|_2^2.
\end{equation}

\subsubsection{Hard-Vicinal Image Denoising Loss (HVIDL)}
\label{sec:hvidl-ctd}

To rigorously address the challenge of data sparsity in the continuous regression label space, we employ the Hard-Vicinal Image Denoising Loss (HVIDL), as introduced in ref \cite{ding2024ccdm}. Training generative models strictly on exact label matches $(\bm{x}, \bm{y})$ is often inefficient because the underlying physical manifold varies smoothly, yet exact samples for any specific parameter $y$ may be rare or non-existent.

HVIDL regularizes the training by ``borrowing'' information from a local neighborhood defined by a radius $\kappa$. We define a perturbed label $\bm{y}' = \bm{y} + \bm{\delta}$, where the perturbation follows $\bm{\delta} \sim \mathcal{N}(\mathbf{0}, \sigma_\delta^2 \bm{I})$. The validity of this neighbor is gated by a hard vicinal weight function:
\begin{equation}
    W_{\mathrm{h}}(\bm{y}, \bm{y}') = \mathbf{1}\{\|\bm{y} - \bm{y}'\|_{2} \le \kappa\}.
\end{equation}
The loss function applies this mask to the whitened diffusion objective derived above. By substituting the conditioning label with the vicinal label $\bm{y}'$, the total objective becomes:
\begin{equation}
  \mathcal L_{\mathrm{HVIDL}}
  = \frac{1}{2} \mathbb E_{\substack{t\sim\mathcal U(0,1)\\\bm \epsilon\sim\mathcal N(\mathbf 0,\bm I)\\\boldsymbol\delta\sim\mathcal N(\mathbf 0,\sigma_\delta^{2}\bm I)}}\Bigl[
      \gamma_{\eta}'(t)\, W_{\mathrm h}(\bm y,\bm y')\,
      \bigl\|\,\bm H_{\bm y'}^{-1/2}\bigl(\bm \epsilon_{\phi}(\bm x_t,t,\bm y', \bm z_0^{(c)})-\bm \epsilon\bigr)\bigr\|_{2}^{2}
    \Bigr].
\end{equation}

Relative to standard pointwise training and soft vicinal weighting schemes, this formulation offers two theoretical advantages. First, it enforces local smoothness and data efficiency. Since neighboring labels typically correspond to perceptually and physically similar data states, HVIDL implicitly augments the effective data density. It compels the network's noise prediction to change smoothly within the local neighborhood of $\bm{y}$ instead of memorizing isolated points.

Second, unlike soft weighting schemes (e.g., Gaussian weights), the hard vicinity constraint acts as a strict gating mechanism. This is crucial for scientific domains where precision is paramount. It ensures that the model effectively ignores ``distant'' physical parameters where $\|\bm{\delta}\|_2 > \kappa$. These distant points, while statistically reachable, often correspond to fundamentally different physical processes, and learning from them would introduce label noise. Furthermore, the quadratic form of the whitened residual preserves the closed-form Gaussian algebra of the VDM framework, which ensures that the optimization landscape remains tractable.

\subsection{Discrete Circuit-Token Decoder}
\label{sec:ctd-decoder}

The Conditional Circuit-Token Diffusion (CTD) produces a clean, continuous embedding $\bm{x}_{0} \in \mathbb{R}^{L \times d_{\mathrm{tok}}}$. While this representation captures the rich semantic relationships between circuit elements, it serves as a continuous relaxation of the target discrete gate sequence. To recover valid quantum operations, we must project this relaxation back onto the discrete manifold.

Decoding entails a single linear projection $f_{\text{dec}}: \mathbb{R}^{d_{\text{tok}}} \to \mathbb{R}^{|V|}$, where $|V|$ is the size of the discrete gate vocabulary. For the $i$-th row of the generated tensor, denoted $\bm{x}_{0,i}$, the probability distribution over possible tokens is computed via:
\begin{equation}
    P(\text{token}_i = v) = \operatorname{Softmax}(\bm{W}_{\text{dec}} \bm{x}_{0,i} + \bm{b}_{\text{dec}}).
\end{equation}
The final discrete circuit is obtained via an argmax step (for deterministic reconstruction) or categorical sampling (for stochastic exploration), as detailed in \cite{furrutter2024quantum}. This step effectively collapses the generated continuous manifold onto the valid quantum gates.

\section{Exemplary Applications}
\label{sec:applications}
The QMLC framework introduced in this article enables the inference of a shared latent representation directly from partially informative observations.
It finds applications in bridging systems identification with generative modeling.
As exemplary applications of this general recipe, this section discusses three applications in learning hidden dynamics in physics and engineering domains where full-information labels are unattainable.

\subsection{Gaussian Boson Sampling}

A learned context space that conditionally generates Gaussian Boson Sampling (GBS) configurations based on desired measurement statistics can be a promising direction for both forward modeling and inverse design in quantum photonics ~\cite{bourassa2021blueprint,madsen2022quantum}. 
Traditional GBS setups rely on fixed interferometers and squeezed input states, which makes the mapping from physical parameters to output photon number statistics analytically tractable but computationally intensive to invert.
Using the proposed generative model trained over the context space of GBS configurations, such as squeezing parameters, interferometer unitaries, and detection patterns, conditioned on target measurement distributions, we can effectively learn an implicit inverse map that bypasses direct simulation.
This approach can be particularly useful for applications in molecular vibronic/energy spectra generation, where the underlying data is inherently quantum. 

\subsection{Machine Learning Quantum Control}

An emerging application of machine learning is quantum control.
Using the proposed framework, traditional circuit compilation can be bypassed, and pulse-level control signals can be synthesized directly, tailored to the desired measurement outcome ~\cite{khaneja2005optimal,fauquenot2025open}. 
This approach of driving the physical qubits directly to produce the intended output distribution is particularly advantageous for near-term quantum devices, where noise characteristics that drift significantly affect overall fidelity. 
By replacing discrete gate decomposition with continuous control synthesis, such techniques promise faster execution, lower overhead, and improved robustness to device variability, ultimately bridging the gap between abstract quantum algorithms and their hardware realization.

\subsection{Quantum Program Synthesis}

This framework also enables applications of empirical quantum algorithmic information theory, shifting from abstract Kolmogorov complexity to learned representations of quantum programs grounded in experimental datasets. 
The shared latent space inferred directly from GST observations reframes quantum program synthesis as a problem of navigating a generative concept manifold, which makes the tradeoff between explainability and efficiency explicit. 
On one end, compressed latent information enables the efficient optimization of performant circuits, aligning with directions such as Quantum Architecture Search for parametric ans\"atze~\cite{xie2025deqompile}, QISA for minimal instruction sets, quantum agency models~\cite{sarkar2022qksa}, and group-invariant geometric QML~\cite{meyer2023exploiting}. Meanwhile, breaking abstractions and co-design~\cite{sarkar2024automated} allows for the optimization of full-stack quantum resources.
This suggests a principled approach to studying how algorithmic complexity, physical realizability, and semantic transparency interact in quantum computing systems.

\subsection{Neuromorphic Hardware Acceleration}
Another promising direction for practical deployment of the proposed QMLC framework is the implementation of its learning and inference pipeline on neuromorphic computing substrates. Neuromorphic processors implement massively parallel, event-driven computation and support spiking or analog neural networks with extremely high energy efficiency, making them attractive platforms for accelerating machine-learning workloads derived from experimental data streams. Recent neuromorphic architectures such as IBM TrueNorth~\cite{merolla2014million}, Intel Loihi~\cite{davies2018loihi}, and mixed-signal memristive systems~\cite{kuzum2013synaptic} demonstrate the ability to perform large-scale neural inference with orders-of-magnitude improvements in energy efficiency compared to conventional von Neumann processors.
In the proposed architecture, the permutation-invariant Set-ViT encoder aggregates collections of GST circuits into a compact latent context representation that captures correlated hardware behavior such as drift, crosstalk, and coherent error accumulation. Mapping this encoder onto neuromorphic hardware would enable highly parallel processing of circuit tokens and measurement embeddings, allowing continuous assimilation of experimental data streams. Similarly, the diffusion-based generative decoder can be implemented using neuromorphic inference engines that efficiently approximate iterative denoising dynamics.
Such an implementation would enable a tightly integrated closed-loop characterization and synthesis system, in which experimental GST data are processed on dedicated neuromorphic accelerators colocated with the quantum control stack. The learned latent context could then be updated in near real time and used to generate hardware-native circuits that adapt to device drift or calibration changes. This hardware-algorithm co-design may significantly reduce classical computational overhead while improving responsiveness of the compilation layer for near-term quantum processors\cite{indiveri2015memory, hadaeghi2021neuromorphic, schmidhuber2015deep}.

\section{Conclusion}
\label{sec:conclusion}
In this article, we propose a new paradigm for quantum circuit synthesis that fundamentally rethinks the interface between hardware characterization and compilation. 
The proposed quantum machine learning control (QMLC) framework bridges three traditionally disjoint layers of the quantum stack: system identification, machine-learning-based control, and program synthesis. 
Instead of treating gate-set tomography (GST) as a preliminary calibration step followed by an idealized unitary decomposition (UD), we introduce an end-to-end generative framework that learns directly from tomographic data and synthesizes circuits whose operational behavior is natively grounded in the device's physics. 
This bypasses the reliance on explicit noise models and rigid physicality constraints, offering instead a fully data-driven route from characterization to construction.

We embed unordered GST datasets into a permutation-invariant latent concept space and couple this representation with a hierarchical diffusion model, which enables robust conditional circuit generation.
The proposed Set Vision Transformer captures global device-level regularities from partially informative observations, while the conditional diffusion decoder translates user-specified measurement objectives, supplied as probability distributions, into executable circuits that respect the empirically learned dynamics of the hardware. 

QMLC opens a pathway toward continual, context-aware compilation in the presence of drift, crosstalk, and calibration challenges of NISQ hardware. 
Beyond circuit synthesis, we discuss how the same conceptual workflow extends naturally to pulse-level control, inverse design in photonic platforms, and empirical approaches to quantum program complexity.
Large-scale implementation and experimental validation of the proposed architecture would be imperative to quantify the practical advantages of generative, tomography-conditioned compilation over conventional pipelines.

\bibliographystyle{unsrt}
\bibliography{references}

\end{document}